\documentclass[conference]{IEEEtran}

\usepackage{fancyhdr}
\usepackage[normalem]{ulem}
\usepackage[hyphens]{url}
\usepackage[sort,nocompress]{cite}
\usepackage[final]{microtype}
\usepackage{flushend}

\usepackage{amsmath,amssymb,amsfonts}
\usepackage{graphicx}
\usepackage{textcomp}
\usepackage{xcolor}
\usepackage{setspace}
\usepackage{amsmath}
\usepackage{algorithm}
\usepackage[noend]{algpseudocode}
\usepackage[T1]{fontenc}
\usepackage{float}%
\usepackage{color}
\newcommand{\ignore}[1]{}
\newcommand{\scheme}{\textsc{CHiRP}}

\usepackage{paralist}
  {\colorbox{yellow}}%
  {}

\usepackage[compact]{titlesec}

\titlespacing{\subsubsection}{0pt}{0ex}{0ex}

\def\BibTeX{{\rm B\kern-.05em{\sc i\kern-.025em b}\kern-.08em
    T\kern-.1667em\lower.7ex\hbox{E}\kern-.125emX}}

\newtheorem{observation}{\textbf{Observation}}

\makeatletter
\newcommand{\linebreakand}{%
  \end{@IEEEauthorhalign}
  \hfill\mbox{}\par
  \mbox{}\hfill\begin{@IEEEauthorhalign}
}
\makeatother

\begin{document}
\onecolumn
\thispagestyle{empty}
\vspace*{1cm}
{\LARGE\bfseries CHiRP: Control-Flow History Reuse Prediction\par}
\vspace{1cm}
{\Large Author manuscript\par}
\vspace{0.6cm}
2020 53rd Annual IEEE/ACM International Symposium on Microarchitecture (MICRO), pp. 131-145 (2020)\par
\vspace{0.4cm}
DOI: \url{https://doi.org/10.1109/MICRO50266.2020.00023}\par
\vfill
{\small \textcopyright\ 2020 IEEE. Personal use of this material is permitted. Permission from IEEE must be obtained for all other uses, in any current or future media, including reprinting/republishing this material for advertising or promotional purposes, creating new collective works, for resale or redistribution to servers or lists, or reuse of any copyrighted component of this work in other works.\par}
\clearpage
\twocolumn
\setcounter{page}{1}

\title{CHiRP: Control-Flow History Reuse Prediction}

\author{\IEEEauthorblockN{Samira Mirbagher-Ajorpaz}
\IEEEauthorblockA{\textit{Computer Science and Engineering} \\
\textit{Texas A\&M University}\\
College Station, USA \\
samiramir@tamu.edu}
\and
\IEEEauthorblockN{Elba Garza}
\IEEEauthorblockA{\textit{Computer Science and Engineering} \\
\textit{Texas A\&M University}\\
College Station, USA \\
elba@tamu.edu}
\linebreakand %

\IEEEauthorblockN{Gilles Pokam}
\IEEEauthorblockA{\textit{Intel Labs} \\
Santa Clara, USA \\
gilles.a.pokam@intel.com}
\and
\IEEEauthorblockN{Daniel A. Jim\'enez }
\IEEEauthorblockA{\textit{Computer Science and Engineering} \\
\textit{Texas A\&M University}\\
College Station, USA \\
djimenez@acm.org}
}

\maketitle

\begin{spacing}{0.99}

  \begin{abstract} 

Translation Lookaside Buffers (TLBs) play a critical role in
hardware-supported memory virtualization. To speed up address translation and
reduce costly page table walks, TLBs cache a small number of recently-used
virtual-to-physical address translations.  TLBs must make the best use
of their limited capacities. Thus, TLB entries with low potential for
reuse should be replaced by more useful entries. This paper contributes
to an aspect of TLB management that has received little attention in
the literature: replacement policy.  We show how predictive replacement
policies can be tailored toward TLBs to reduce miss rates and improve
overall performance.

We begin by applying recently proposed predictive cache replacement 
policies to the TLB. We show these policies do not work well 
without considering specific TLB behavior.
Next, we introduce a novel TLB-focused predictive policy, Control-flow
History Reuse Prediction (\scheme). This policy uses a history
signature and replacement algorithm that correlates to known
TLB behavior, outperforming other policies.

For a 1024-entry 8-way set-associative L2 TLB with a 4KB page size, we
show that \scheme{} reduces misses per 1000 instructions (MPKI) by an
average 28.21\% over the least-recently-used (LRU) policy, outperforming
Static Re-reference Interval Prediction (SRRIP)~\cite{jaleel2010high},
Global History Reuse Policy (GHRP)~\cite{Mirbagher2018ISCA} and
SHiP~\cite{SamiraMICRO}, which reduce MPKI by an average
of 10.36\%, 9.03\% and 0.88\%, respectively.

\end{abstract}
  \section{Introduction} \label{sec:intro}

Virtual-to-physical address translation is expensive~\cite{Juan1997Reducing,
Kadayif2002Generating, Kadayif2004Compiler, Fan2005Energy,
Bhargava2008Accelerating,Barr2010Skip, Barr2011SpecTLB, Sodani2011Race,
Li2013PSTLB, Gandhi2014Efficient, Pham2015Large, Karakostas2016Energy}. 
Translation lookaside buffers (TLBs) help minimize the need for 
costly page table walks by caching recently 
retrieved virtual-to-physical address mappings~\cite{Couleur1968Shared, 
Clark1985Vax}.

Recent studies by Google~\cite{GoogleAI2014isca}, asmDB~\cite{asmDBISCA19},
and Facebook~\cite{SoftSKUsISCA19} confirm that modern deeply pipelined
speculative OoO CPUs face increasing challenges associated with TLB
performance.  For example, server workloads show growing code footprints
and working set sizes~\cite{Ferdman2012CloudSuite, Kumar2018Shotgun,
Ayers2018Hierarchy, GoogleAI2014isca}, placing tremendous pressure on
caches and TLBs~\cite{IntelWhitePaper19}. The caches and TLBs of future
systems will need to improve at a similar rate to maintain performance.

\begin{figure}[ht!] 
\centering
\includegraphics[width=0.44\textwidth]{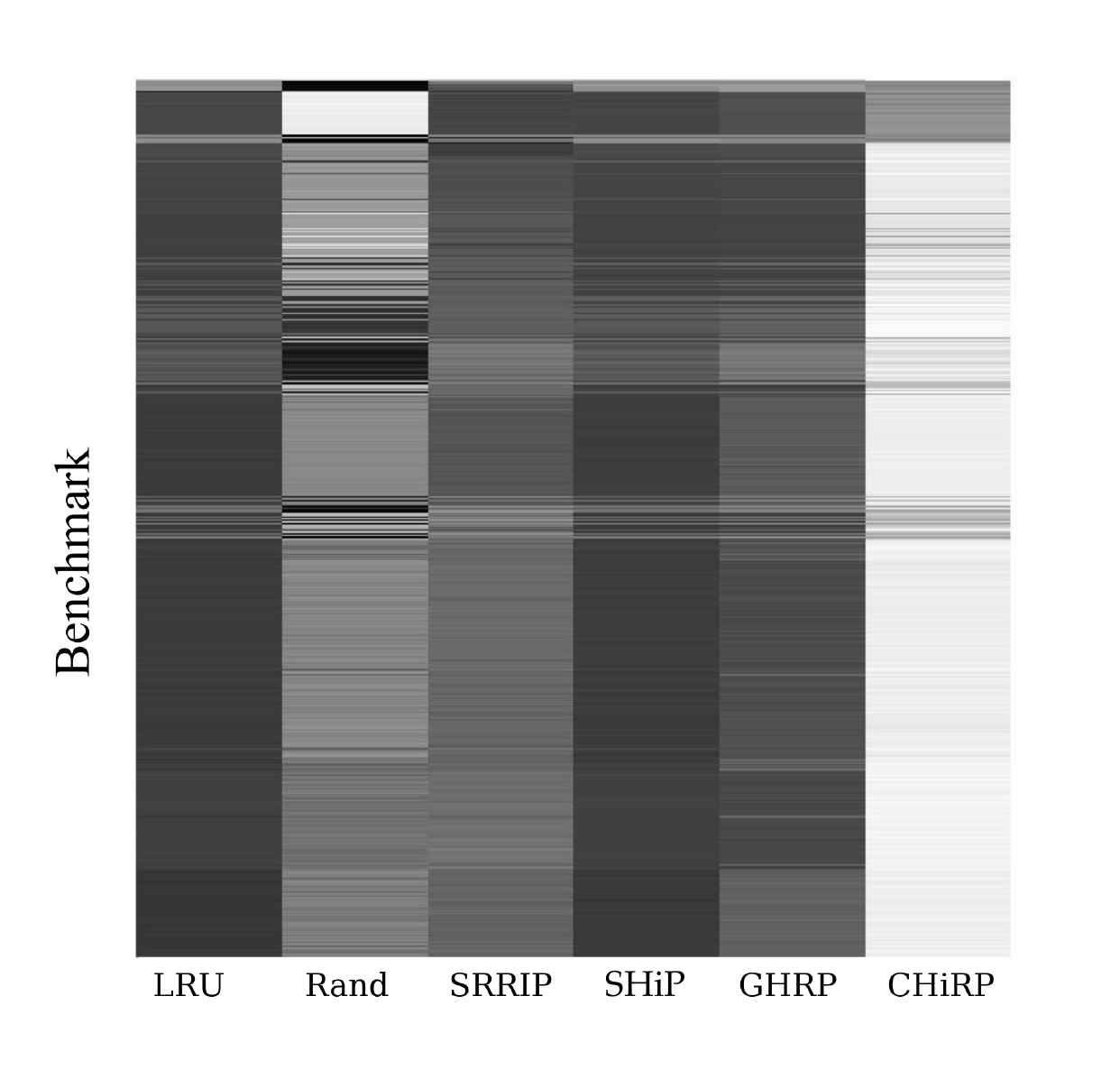}
\caption{Comparing predictive policy efficiency with a heat map shows 
  \scheme{} maintains more live TLB entries 
  compared to other policies when analyzed on 870 different benchmarks.
  A lighter color block indicates higher TLB efficiency, while darker denotes lower
  efficiency.}
\label{fig:heatmap}
\end{figure}

Unfortunately, TLBs are limited in size, and thus reach, due to power,
timing, and area constraints~\cite{Jacob2007Memory}. The TLB lies on
the critical path to accessing memory. Thus, 
{\em increasing L2 TLB sizes to
reduce TLB misses is difficult} because larger TLBs incur higher access
latencies~\cite{Ryoo2017Rethinking}.

Meanwhile, TLB misses are a first-order concern in terms
of their negative impact on performance. Recently studies{{~\cite{gras_translation_2018,CuckooISCA2020,babelfishISCA2020}}}
indicate that many programs can spend hundreds of extra
cycles conducting address translations that did not hit in the TLBs. This is 
despite the fact that the Skylake architecture includes special MMU/paging structure 
caches (or PSCs) to lessen the page walk penalty~\cite{Intel2016Architectures}.
{The study}~\cite{gras_translation_2018}{ finds L2 TLB miss costs range 
from 16.3 cycles for Sandy Bridge in 2011, increasing up to 212 
for Skylake in 2015, 272 cycles for Broadwell Xeon in 2016, and 230 cycles 
for Coffee Lake in 2017.  Such overhead is likely to be exacerbated in the future}\footnote{The new generation of Intel processors, Sunny Cove~\cite{Intel2018}, introduced 
 5-levels radix page tabling.}{given that modern computing platforms can now be supplied with terabytes, and even petabytes, of main memory}
{~\cite{hady2017platform,qureshi2011phase}}
{, all on various memory-intensive workloads that are rapidly emerging}{~\cite{CuckooISCA2020, GoogleAI2014isca,asmDBISCA19,SoftSKUsISCA19}.}

Translation
overheads running into 100+ cycles have also been reported in prior
work~\cite{Pham2015Large, Gandhi2014Efficient}.  Address translation
latencies due to TLB misses represent between 20\% and 50\% of system
run-times today~\cite{Bhattacharjee2017Triggered, Yan2017Coherence,
Bhattacharjee2010Cooperative, Cox2017Multiple, Pham2015Large,
Bhattacharjee2013Large, Bhattacharjee2011Shared, Barr2011SpecTLB,
Karakostas2015Redundant,Basu2013Efficient,Gandhi2014Efficient, Barr2010Skip}
and consume a substantial share of processor energy~\cite{Juan1997Reducing,
Kadayif2002Generating, Kadayif2004Compiler, Fan2005Energy,
Papadopoulou2015Prediction, Karakostas2016Energy, Sodani2011Race}.

Peng {\em et al.} conduct a
thorough study of TLB behavior of Java applications ~\cite{Peng2006JavaApps},
reporting 230+ cycle TLB miss latencies and indicate TLB miss overhead
accounts for 5.5\% to 19\% of the total execution time.
Their study finds that five out of seven benchmarks exhibit similar TLB
overhead.

These concerns motivate us to investigate mechanisms to improve TLB
performance that do not require increasing TLB sizes. Similar
efforts to improve TLB performance have included using varied page
sizes and superpages~\cite{Talluri1992Tradeoffs, Talluri1994Surpassing,
Navarro2002Practical, Ganapathy1998General, Swanson1998Increasing,
IntelWhitePaper19} as well as prefetching~\cite{Saulsbury2000Recency,
Kandiraju2002Going, Bhattacharjee2009Characterizing,
Bhattacharjee2010Cooperative}.

Fortunately, TLBs' organization makes them amenable to predictive
replacement policies.  TLBs are organized with tagged set-associative
SRAM arrays much like cache memories.  Predictive replacement policies
have been well-explored and have been shown to perform well in data
caches~\cite{Kaxiras07,SamiraMICRO,SHiP,PerceptronReuse} that depend on
spatial and temporal locality of data accesses to maintain useful entries.
Access patterns to TLBs are similar to cache accesses at a
larger granularity.  Thus, it is reasonable to apply previous work on
cache replacement/management to TLBs.

TLB replacement policy has received little attention
in the literature. Recent work~\cite{Bharadwaj2018Scalable,
Bhattacharjee2017Triggered, Cox2017Multiple, Pham2014Increasing,
Pham2012CoLT, Parasar2018Seesaw, Pham2015Large, Bhattacharjee2013Large,
Bhattacharjee2010Cooperative} advocates using an LRU replacement policy for 
all levels of TLBs. Other prior work focuses either on reducing the 
cost of a page table walk upon a TLB miss~\cite{Saulsbury2000Recency,
Kandiraju2002Going, Bhattacharjee2009Characterizing,
Bhattacharjee2017Triggered, Barr2011SpecTLB} or reducing the TLB miss 
rate by extending the size of the TLB~\cite{Ryoo2017Rethinking}. 
In this paper, we suggest tackling the fundamental problem of the TLB's 
insufficient capacity by improving its replacement policy.

Our work builds on prior predictive replacement policies geared
toward the last-level cache (LLC), such as static re-reference 
interval prediction (SRRIP)~\cite{jaleel2010high}, signature-based hit 
prediction (SHiP)~\cite{SHiP}, and Global History Reuse 
Prediction (GHRP)~\cite{Mirbagher2018ISCA}, to extract key insights 
for the TLB. We propose a novel mechanism, Control-flow History Reuse 
Prediction (\scheme), that provides superior prediction accuracy and 
performance by better correlating to TLB reuse behavior.

We begin with predictive policies adapted from the cache replacement
literature, in particular the last-level cache (LLC), and show that 
they are not a good fit for TLBs.  We show that features used by these 
schemes do not correlate well to TLB reuse, resulting in negligible 
performance gains. Moreover, LLC-focused prediction policies are 
designed with less stringent cycle time requirements and can 
tolerate several accesses to their prediction tables. TLBs, on the 
other hand, have tighter timing requirements for TLB access. 
Based on this and other insights, we introduce a policy that 
efficiently indexes prediction tables using a novel signature 
specifically designed to correlate to TLB behavior. We focus on 
the L2 TLB as L2 TLB misses account for most of the cycles spent
in the TLB miss handler~\cite{Basu2013Efficient}.

This paper makes the following contributions:

\begin{enumerate}

\item A first study and exploration of TLB replacement policies by
implementing and adapting policies from previous work on data caches 
and branch target buffers to the TLB.

\item An intuition on why previous predictive replacement
policies may or may not be as effective on TLBs. We evaluate the 
impact of various optimizations on adapted predictive replacement 
policies over a large suite of industry-sourced traces.

\item A new predictive replacement policy, Control-flow History Reuse
Prediction (\scheme). This policy indexes prediction tables using a
signature specially designed to correlate with TLB behavior. It reduces 
L2 TLB misses by 28.21\% on average over LRU, resulting in significant speedup.
For example, for a page walk latency of 150 cycles, \scheme{} yields a 
geometric mean speedup of 4.8\%. 

\end{enumerate}
  \section{Background}

Processor performance is affected by the TLB in two ways: the \textit{number
of TLB misses} and the \textit{TLB miss penalty} in cycles.  While other
solutions have mainly focused on reducing the TLB miss penalty, very
little work has focused on reducing the number of misses in the TLB
directly.  There have been a handful of papers on prefetching into the
TLB~\cite{Saulsbury2000Recency,Kandiraju2002Going,Bhattacharjee2010Cooperative}.
However, to the best of our knowledge, no previous work has
proposed a predictive replacement policy specifically for
TLB. Rather, recent work employs LRU or Random replacement
policies~\cite{AbhishekSynthesis,Bharadwaj2018Scalable,
Bhattacharjee2017Triggered, Cox2017Multiple, Pham2014Increasing,
Pham2012CoLT, Parasar2018Seesaw, Pham2015Large, Bhattacharjee2013Large,
Bhattacharjee2010Cooperative}. We advocate using a predictive replacement
policy that relies on a variety of program features to guide TLB entry
replacement to improve performance without needing to increase the TLB's
size.

Recent work in cache and BTB replacement shows that reuse prediction can
significantly reduce misses and improve
performance~\cite{Lai2001,Liu2008,SamiraMICRO,SHiP,PerceptronReuse,Hawkeye,
MultiperspectiveReuse,Mirbagher2018ISCA}.
Predictive replacement policies attempt to predict whether a cached item will
be used again before it is evicted.  If not, then it is a prime candidate for
eviction. This idea is superior to LRU replacement, in which a block with no
near-term reuse must migrate all the way down the recency stack before being
replaced.  However, a highly accurate predictive replacement policy for one
cache-like structure may not work for another cache-like structure.  For
example Mirbagher{\em et al.}~\cite{Mirbagher2018ISCA} show that while
PC-based policies such as SDBP~\cite{SamiraMICRO} and SHiP~\cite{SHiP} reduce
the number of dead blocks in the LLC, it is detrimental to instruction caches
and BTBs. We find the same applies to TLBs.

There are three main challenges in designing a predictive replacement
policy. The first is finding the microarchitectural features that correlate
with reuse for a particular cache-like structure. These features vary
for different structures such as the TLB and data caches, and even different
applications~\cite{Lai2001,Liu2008,SamiraMICRO,SHiP,PerceptronReuse,
MultiperspectiveReuse,Mirbagher2018ISCA}.
The second is building an efficient {\em signature} by combining the
identified correlating features. The features are combined to reduce their
hardware storage budget and prediction time.  The third is designing a
fast/low-cost prediction algorithm to use this signature. The latter is
particularly important for the TLB as it lies on the critical path to a
memory access.

Once we identified highly correlating features of TLB entry reuse, we adapted
previous algorithms to propose a novel, low-cost algorithm specifically
tailored for reuse prediction in L2 TLBs.  Previous work on LLC reuse
prediction that uses prediction tables has used multiple features hashed to
multiple indices~\cite{SamiraMICRO} or
signature~\cite{PerceptronReuse,MultiperspectiveReuse} to combine several
predictions into one.  Because the TLB is on the critical path to accessing
memory, we reduce accesses to a {\em single table} with a signature 
combining several features as the most latency-sensitive approach.

We explore using predictive cache replacement policies such as static
re-reference interval prediction (SRRIP)~\cite{jaleel2010high},
signature-based hit prediction (SHiP)~\cite{SHiP}, and Global
History Reuse Prediction (GHRP) \cite{Mirbagher2018ISCA} for the TLB,
and propose a new mechanism, Control-flow History Reuse Prediction
(\scheme), to better guide TLB entry replacement. 

\subsection{Static Re-Reference Interval Prediction}

SRRIP~\cite{jaleel2010high} predicts which blocks will be referenced
again ({\em i.e.} re-referenced) in the cache.  Each block has a 2-bit re-reference 
prediction value (RRPV) placing the block into one of four categories ranging from
near-immediate re-reference to distant re-reference.  A first prediction is
made on block placement and revised when a block is reused or replaced. Blocks 
with distant re-reference prediction are evicted.  If
there are none, the RRPV for each block in the set is incremented until there
is at least one eviction candidate. We adapt SRRIP to work with TLB entries
instead of cache blocks.

\subsection{PC-Based Dead Block Predictors}

In sampling-based Dead Block Prediction (SDBP)~\cite{SamiraMICRO}, a predictor learns 
the pattern of accesses and evictions from a small number of sets kept in a structure 
called the sampler. When a load or store accesses the LLC, the address (PC) of that 
instruction is hashed to index prediction tables. Counters read from the tables 
are summed and thresholded to predict whether the block is dead. In the original 
SDBP paper, blocks are predicted on each access\cite{SamiraMICRO}. Signature-based 
Hit Prediction (SHiP) improves on this idea by using the prediction only for placement 
in a RRIP-replaced cache, reducing the number of predictions and significantly 
improving performance.

However, sampling is not suitable for structures indexed by instruction addresses 
such as the BTB and instruction cache~\cite{Mirbagher2018ISCA}. Sampling works 
for data caches because the behavior of a memory access instruction, represented 
by its PC, generalizes over the entire cache. Instruction streams do not allow 
set sampling to generalize the behavior of accesses to such structures since 
the PC itself forms the index into the structure.  

We find that sampling also does not work well for second-level TLBs. The reason is 
the coarser granularity of TLB entries versus cache blocks. A PC accesses different 
data addresses that are in the L2 TLB, which might lead one to believe sampling should 
generalize across the TLB. However, in the LLC, one sampled set may map to many cache sets all accessed by the same PC, which allows behavior to be generalized across sets. 
On the other hand, in the L2 TLB, one PC accesses data that are mapped to much fewer 
TLB entries than cache blocks. Spatial locality for data accessed by a single PC 
does not expand beyond a few TLB entries, so generalization fails.

Because of this failure, in this work we evaluate SHiP with the same general algorithm,
but with bits of PC kept as metadata in each TLB entry, which is equivalent to keeping 
a sampler the same size as the structure. We consider SHiP to be the best 
cache replacement policy from previous work that would be implementable under 
the tight timing requirements of the TLB access critical path.

\subsection{Global History Reuse Prediction}

Global History Reuse Prediction (GHRP)~\cite{Mirbagher2018ISCA} is
the state-of-the-art predictive replacement policy for BTB and i-cache
replacement. We adapt GHRP for TLB replacement. GHRP has a structure
similar to SHiP, but the signature used to index the prediction tables is
specifically designed for instruction streams. Like a branch predictor,
it uses the global history of conditional branch outcomes~\cite{gshare}
as well as lower-order bits from branch addresses to form an index into
a table of counters that keep track of reuse behavior.

\subsection{Offline Learning}
We use insights from neural networks to design a new hand-crafted feature
that represents a program's control-flow history compactly and that can be
used with a much simpler linear learning model. Offline training has been
used for designing replacement policies in the past through using genetic
algorithm by Jim\'enez {\em et al.}~\cite{Jimenez2017MultiperspectiveRP} and
LSTM by Shi {\em et al.}~\cite{Glider2019}. Their work shows how insights
from offline training can improve learning model for online prediction in the
LLC. We use {\sc Adaline} (ADAptive LINear Element)~\cite{WidrowHoff60,Widrow90} 
to find insights for TLB replacement policy.

{\sc Adaline} uses a vector of weights that records correlations between an 
input vector and a target value. It can be used to classify inputs into one of two
classes. 

{\sc Adaline} computes the weighted sum of the input patterns $x(n)$.
\[
y(n)=w^T(n)x(n)+\theta
\]
{\sc Adaline} weights are updated after the desired outcome $d(n)$ of the predicted event is known. 
If the prediction was correct then the weights remain unchanged. Otherwise, the inputs are used to update the corresponding weights.
\[w(n+1)=w(n)+\mu[d(n)-y(n)]x(n)\]
where $\mu$ is the learning-rate parameter and the difference $d(n)
-y(n)$ is the error signal. 

\subsection{\scheme{}}
We explored adapting predictive cache replacement policies to the TLB and
observed that features that correlate well to cache reuse behavior may not
necessarily correlate well to TLB reuse behavior.  In contrast to a
cache access, a TLB access is of coarser granularity with many PCs
that map to the same TLB entry.  Furthermore, depending on the
context, each such PC may result in an eviction or a reuse of the same
TLB entry. We find that predicting a TLB entry's reuse requires multiple 
features that we compose into a single signature for better prediction accuracy 
and overhead reduction.  

  \section{The Reuse Prediction Problem in TLB \& Our Solution}

We find that predictive policies for the LLC, instruction cache, and BTB do not 
apply well to L2 TLBs, and describe the main reasons why in this section. 

We simulated 870 workloads from a variety of categories provided publicly by Qualcomm~\cite{CVP1website} to prevent overfitting to one type of workload. 
More information about the full details of our simulation methodology can be
found in Section~\ref{sec:method}.

We first applied signature-based hit prediction
(SHiP)\cite{SHiP}, which was shown to be useful in the LLC.
SHiP uses only the address (PC) of the most recent instruction.
However, our results show that a solely PC-based
reuse entry prediction does not perform much better than LRU, giving a
reduction in MPKI of only 0.88\%.

We investigated whether aliasing was the cause of the observed mispredictions, but 
found that even with an unlimited prediction table size (i.e. no aliasing), SHiP is 
not able to detect dead entries in the TLB, giving a reduction in MPKI of only 0.63\%. 
Since prediction table size was not the source of the mispredictions, we further 
investigated by limiting the prediction to only a subset of the TLB sets and used 
LRU for the rest. This technique also just slightly improves accuracy, reducing 
MPKI by 1.28\%, leading to the following observation:

\begin{observation}
  \textit{The inaccuracy in previous predictive policies for the TLB is not 
    due to conflicts among multiple sets but rather within the sets themselves.}
\end{observation}

We find that a TLB entry may experience many hits from one or more PCs that map to 
the same entry before it is eventually evicted. This is because a larger range 
of unique addresses map to the same entry in the TLB compared to accesses to a block 
in a cache. Indeed, there is a nearly two order-of-magnitude difference between a 
4KB page and a 64B block.

Therefore, we obtain our second observation:

\begin{observation}
  \textit{The coarse-grained nature of TLB accesses results in increased aliasing 
    in previous predictive policies, which cause the prediction counters to 
    saturate too quickly, rendering the predictor ineffective.}
\end{observation}

From Observation 2 we posit that in order to dissipate this noise, we need to slow down the 
rate at which the prediction counters are updated. We do this by limiting updates 
{only to hits of a TLB set different from the one last accessed.}
We call this method { \em Selective Hit Update}. 
Selective Hit Update improves accuracy by reducing average MPKI by 5.85\%.

\begin{figure}[htb!] 
\centering
\includegraphics[width=\linewidth]{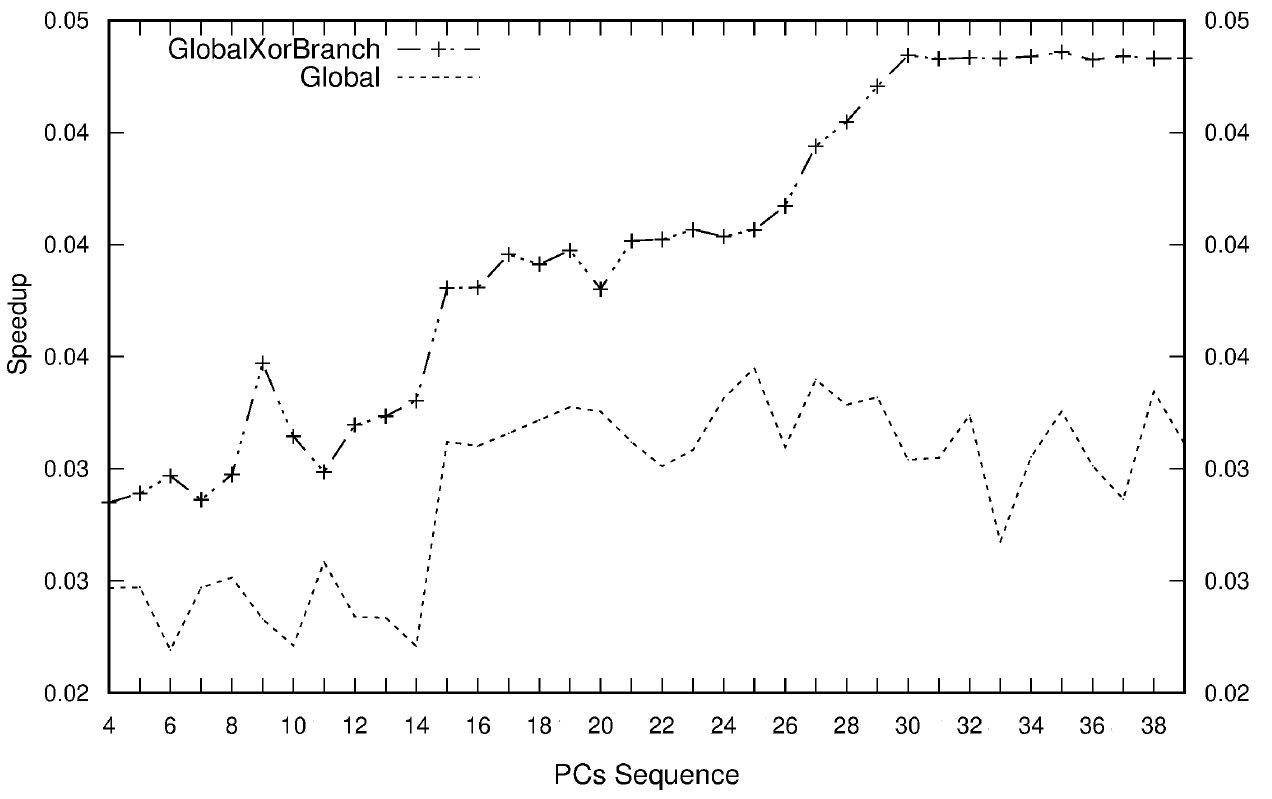}
\caption{Speedup does not increase for global PC history length more
  than 15. However, combining branch history into signature, \scheme{}
  can benefit from history lengths longer than 30.}
\label{fig:historylength}
\end{figure}

Previous work~\cite{Mirbagher2018ISCA ,Glider2019} has shown that
a longer history of past PCs would benefit predictive replacement
policies in the LLC and i-cache. Figure~\ref{fig:historylength} shows our
results conducting a similar study for the TLB. Here, we analyze varying PC 
history lengths from 4 to 40 and their resulting speedups. We find that the 
benefits of using longer global PC history for TLB reuse prediction diminishes 
beyond a length of 15. This contrasts with prior work on predictive policies for 
the LLC, which show benefits of using global PC history length of 60 or more. This 
is likely due to the coarse-grain nature of TLB accesses that may limit the global 
history window from capturing enough information pertaining to TLB reuse. To improve 
on this, we augment the global PC history with branch path history information, 
resulting in a history length greater than 30 (Figure~\ref{fig:historylength}). 
Hence, our third observation is as follows:

\begin{observation}
  \textit{TLB reuse prediction does not benefit from
    a global PC history of length 15 or more. However, by combining
    branch path history into a prediction signature, \scheme{} can take 
    advantage of a PC history length of 30 or more.}
\end{observation}

Branch history is effective because L2 TLB accesses come from both data and 
instructions in the first-level TLBs. Conditional branch histories can reflect 
the data accesses when global path history does not. Branch path history can 
also reveal high-level program semantics that also contribute to TLB misses.

\subsection{PC Bits Carry Uneven Weights}\label{sec:adeline}

Previous work~\cite{Jimenez2017MultiperspectiveRP, Glider2019} 
shows that certain features from program behavior are important to predicting reuse
of a block in the LLC. We come to the same conclusion with
regard to TLBs, recognizing that some bits of the PC carry more weight than
others in reuse prediction. To show this for the case of TLBs, we use 
the weights of a trained {\sc Adaline} neural network to score the bits of PCs 
that we incorporate into the global history.  The idea is based on the principle 
that the weights of the input nodes corresponding to less important features are 
expected to be smaller in trained {\sc Adaline} networks. The incorporation of 
appropriate regularization terms in the {\sc Adaline} update function encourages 
such weights to converge to zero and ultimately be eliminated. 

\begin{figure}
\centering
\includegraphics[width=3.9in]{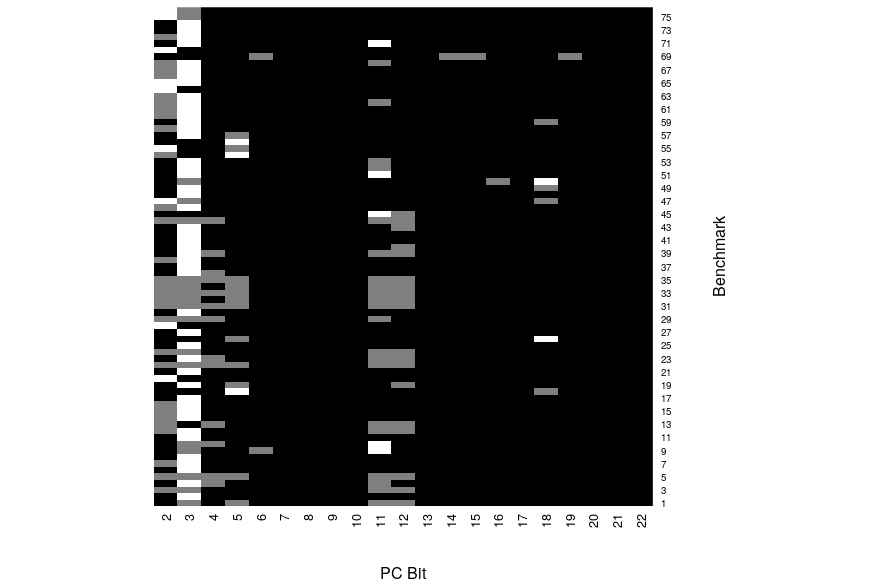}
\caption{Each row represents an offline-trained {\sc Adaline} weight vector 
  for one benchmark. The $x$-axis shows the PC bit used as input. The white 
  boxes show reuse prediction in TLB entries is strongly correlated with bits 
  2 and bit 3 of the PC.}
\label{fig:bits}
\end{figure}

Figure~\ref{fig:bits} shows that the two lower-order bits of a PC address (bits 2 and 3) 
contain important information, as indicated by their higher-weight color values. Thus, 
passing on these bits to the signature function yields a high chance of preserving 
information to reduce aliasing. In our proposed \scheme{} policy described in 
Section~\ref{sec:CHiRPalg}, we keep these two correlated bits in the global path 
history.

\subsection{Modeling Efficient Signatures}\label{sec:XOR}

Aliasing in the prediction table is harder to solve in the TLB than caches. With TLB 
reuse prediction, far too many PCs map to the same TLB entry, {\em i.e.}, 64 times 
more than to a cache block. The problem of aliasing is exacerbated further with 
large footprint applications.

If a counter in the prediction table changes direction frequently due to aliasing, 
the same problem will only be exacerbated with a smaller table size. To achieve 
high reuse prediction accuracy with a smaller table size, we have to solve aliasing 
first. 

This problem can be addressed by coordinating how the input bits are transformed 
by designing a succinct signature. We found that employing shifting and scaling 
techniques as described by Lecun and Hinton \cite{lecun2012efficient, hinton2012neural} 
improves prediction accuracy. 

We accomplished this by injecting and shifting leading zeros into specific bit 
positions of different components of the signature including the global 
path history, conditional branch history, indirect branch history, and the shifted 
PC of the access (section~\ref{sec:CHiRPalg}). 

Doing this both shifts the individual PCs and scales the less salient history bits 
down to make them \textit{less visible} to the learning process, allowing the 
prediction table to converge to an accurate counter value with 3 times fewer 
entries than GHRP. 

The above techniques of shifting and scaling the signature bits are simple to 
implement in hardware and provide significant reduction in TLB MPKI. 
Figure~\ref{fig:motive} shows that while adding conditional branch path history 
to the signature would reduce MPKI by 23.88\%, adding two leading zeros in 
the path history would allow the effect of conditional branch history to 
reduce MPKI by 26.98\%.

In the next section we discuss our signature function and the individual effect of above optimizations.

  \begin{figure*}
\centering
\includegraphics[width=0.78\textwidth]{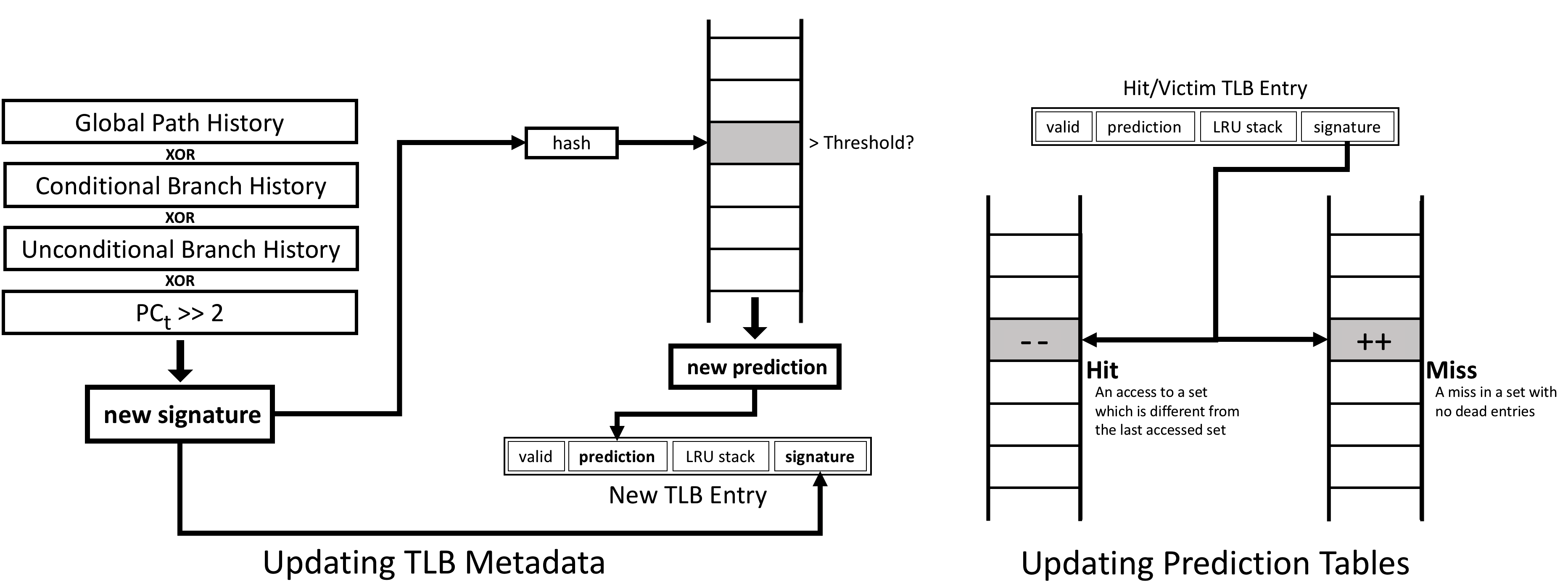}
\caption{\scheme{} TLB metadata and prediction table update flow using a signature. }

\label{fig:algdiagram}
\end{figure*}

\begin{figure*}[ht!]
  \begin{tabular}[t]{@{} p{.49\linewidth} @{}}
    \hline

    \begin{algorithmic}[1]
\begin{small}
      \State int $\textit{predTable}[\textit{numCounters}]$

\Procedure{AccessTLB}{int \textit{VA}}

\State $\textit{set} \gets \text{calcSet\textit{(VA})}$
\State $\textit{isMissed} \gets \text{isTagMatch(\textit{VA})}$
\State $\textit{sign} \gets\textit{VA>>2} \oplus  \textit{pathHist} \oplus \textit{condBrHist}  \oplus \textit{unCondBrHist} $ \label{op:xorhistories}
\State $\textit{index} \gets \text{Hash(\textit{sign})}\mod 2^{16}$ \label{op:indexsignnew}
\State $\textit{cntrNew} \gets \text{predTable[index]}$ \label{op:cntrssignnew}

\If {$\textit{isMissed} = \textit{true} $}\Comment{miss}
\State $\textit{entry} \gets
\text{victimEntry(\textit{set})}$\label{op:choosvict}

\If {\textit{entry.isDead = false}}\Comment{lru} \label{op:lruselected}
\State $\textit{index} \gets \text{Hash(\textit{entry.signature})}$ \label{op:indexssignold}
\State $\text{updatePredTables(\textit{index, true})}$  \label{op:updatedeadevict}

\EndIf
\State{$\textit{entry.firstHit} = \textit{true}$} \Comment{insertion} \label{op:insert} 
      
\Else\Comment{hit} \label{op:hit}
\State $\textit{entry} \gets \text{matchedEntry(\textit{set, tag})} $

\If {$\textit{entry.firstHit} = \textit{true} $}\Comment{access table}  \label{op:newset}

\State $\textit{indices} \gets \text{Hash(\textit{entry.signature})}$ \label{op:indexsignhit}
\State $\text{updatePredTables(\textit{index, false})}$   \label{op:updatealive}
\State $\textit{entry.dead} \gets \text{predict(\textit{cntrNew, deadThresh})}$ 
\State $\textit{entry.firstHit} \gets \textit{false}$ 

\label{op:predict3}
\EndIf

\EndIf
\State $\textit{entry.signature} \gets \textit{sign} $ \label{op:updatsign}
\State $ updateLRUstackPosition() $

\State $ \text{UpdatePathHist(\textit{VA, pathHist})}$ \label{op:buildhistory}

\If {$\textit{instType} = \textit{conditionalBranch} $} \label{op:allRet}
\State $\text{UpdateBrHist(\textit{VA, condBrHist})}$ \label{op:uprethistory}
\EndIf
\If {$\textit{instType} = \textit{unConditionalBranch}$}   \label{op:allDirJump}
\State $\text{UpdateBrHist(\textit{VA, uncondBrHist})}$ \label{op:upjumphistory}
\EndIf
\EndProcedure
\end{small}
\end{algorithmic} \\

  \end{tabular}\hfill
  \begin{tabular}[t]{@{} p{.47\linewidth} @{}}
    \hline
    \begin{algorithmic}[1]
\begin{small}
\setcounter{ALG@line}{27}
      \Procedure{UpdatePathHist}{int \textit{VA}, int \textit{history}} \label{op:historypath}

\State $\textit{history} \gets \textit{history} << 4 $ 
\State $\textit{history} \gets (\textit{history } | \textit{ VA}_{2..3})$
\EndProcedure
\vspace*{5mm}
\Procedure{UpdateBrHist}{int \textit{VA}, int \textit{history}} \label{op:historybranch}
\State $\textit{history} \gets \textit{history} << 8 $ 
\State $\textit{history} \gets (\textit{history } | \textit{ VA}_{4..11})$
\EndProcedure
\vspace*{5mm}

\Procedure{predict}{int \textit{counter}, int \textit{threshold}}
\If {\textit{counter} > \textit{threshold}}
\Return true
\Else
{ \Return \textit{false}}
\EndIf
\EndProcedure
\vspace*{5mm}
\Procedure{victimEntry}{Set \textit{set}} \label{alg:choosvict}

\For { \text{int} \textit{i} = 1 \textit{to associativity}  }
\State $\text{entry} \gets \textit{ set.entries}\text{[\textit{i}]} $
\If {\textit{entry.isDead = true}}
\Return \textit{entry}
\EndIf
\EndFor 

\Return \text{LRUentry()}
\EndProcedure
\vspace*{5mm}

\Procedure{updatePredTable}{int \textit{index, bool Dead}} \label{alg:updatetables}
\If  {\textit{ Dead = true}}
\State $ \textit{ predTable}[\textit{index}]  {++} $
\Else
\State $ \textit{ predTable}[\textit{index}]  {--} $
\EndIf
\EndProcedure
\end{small}

   \end{algorithmic} \\
  \end{tabular}
\vspace*{-5mm}  \caption{\scheme{} algorithm.}
\label{fig:mhrpalgo}
\end{figure*}

\section{Control-flow History Reuse Prediction Algorithm} \label{sec:CHiRPalg}

\subsection{Overview} \label{sec:overview}

\scheme{} correlates TLB replacement with reuse history. \scheme{} uses features 
that best correlate to reuse behavior and combines them into a signature that 
is used to uniquely tag each TLB entry (\ref{ssec:global:signature}). 

This signature is subsequently used to track the reuse behavior of the associated 
TLB entry by means of a prediction table indexed by the signature 
(\ref{ssec:global:predtable}). The prediction table is updated on an eviction or 
a reuse, and the resulting prediction status is written back into the corresponding 
TLB entry to inform the next TLB replacement operation
(\ref{ssec:global:mhrpoperation}). 
Figure~\ref{fig:algdiagram} describes the main components of \scheme{} and
Algorithm~\ref{fig:mhrpalgo} provides the \scheme{} algorithm.

\subsection{\scheme{} Signature} \label{ssec:global:signature}

\scheme{} contributes four features that correlate with reuse behavior.
The first is the global path history of PCs. The global path history
in \scheme{} is 64 bits wide and is updated on each access by shifting
the two lower-order bits of the PC into the path history, followed by
two zero bits (Figure~\ref{fig:mhrpalgo}, line~\ref{op:historypath}),
as previously discussed in subsections \ref{sec:adeline} and
\ref{sec:XOR}, respectively. The global path history in \scheme{} allows
recording the last 16 accesses.

The second and third features are the conditional and unconditional indirect
branch address history, respectively.  Each of these histories is 64 bits
and is updated by shifting the eight bits of the PC [11:4] into the 
branch history on every conditional (resp. unconditional indirect) branch 
instruction (Algorithm~\ref{fig:mhrpalgo}, line~\ref{op:historybranch}.),
recording the last 8 branch accesses for each type.

The fourth feature is the current PC, shifted right by two bits. 
The signature is constructed by XOR-ing the global path history with 
the conditional branch history, the unconditional indirect branch history, 
and the shifted PC of the access (Figure~\ref{fig:mhrpalgo}, 
line~\ref{op:xorhistories}).

To compute indices into the prediction table, \scheme{} computes a 16-bit hash of the constructed
signature. {For hashing, we first use Robert Jenkins' 64-bit mix 
function}{~\cite{Hash}}{. The mix function enables a single-bit 
change in the key to influence widely disparate bits in the hash result. We then take the modulo of the table size to generate the hash table index}(Figure~\ref{fig:mhrpalgo}, line~\ref{op:indexsignnew}).

Note that the signature relies on bits from the branch PC, not conditional
branch outcomes or bits from branch targets.

\subsection{\scheme{} Prediction Table} \label{ssec:global:predtable}

\scheme{} stores metadata for each L2 TLB entry, consisting of 3 LRU stack position
bits, a valid bit, a 16-bit signature and a prediction bit 
(See Figure~\ref{fig:algdiagram}, Updating TLB Metadata). 
\scheme{} uses a table of saturating counters to provide a prediction. 
The table is indexed by a hash function of the signature.  The corresponding 
counter is thresholded, and if the counter exceeds the threshold, the entry is 
predicted as dead.

\subsection{\scheme{} Operations} \label{ssec:global:mhrpoperation}

In contrast to SHiP and GHRP that require updating the prediction table on
each TLB access, the bulk of \scheme{} operations occurs off the TLB critical
path, with minimal impact to TLB latency.  In particular, \scheme{} updates
its prediction table on a TLB miss only if the selected victim is LRU ({\em
i.e.} no dead entry is found).

The operations pertaining to a TLB miss involve (1) selecting a victim, (2)
updating the victim's reuse history in the prediction table if the victim is
LRU, and (3) updating the prediction metadata for the new TLB entry.

\paragraph{Victim selection} On a TLB miss, \scheme{} first attempts to 
select a victim among the entries predicted as dead. If no such entry
is found, \scheme{} evicts the LRU entry (Figure~\ref{fig:mhrpalgo}, line
~\ref{alg:choosvict}).

\paragraph{Prediction table update} Because \scheme{} updates its
prediction table only if the victim is LRU (Figure~\ref{fig:mhrpalgo},
line~\ref{op:lruselected} --~\ref{op:updatedeadevict}), evicting the LRU entry 
effectively makes it a dead candidate the next time around. This justifies why 
the prediction table needs be updated. The signature of the victim entry is used 
to index the prediction table and the corresponding counter is incremented, 
since the entry was just shown to be dead (Figure~\ref{fig:mhrpalgo}, 
line~\ref{alg:updatetables}).

\paragraph{Prediction metadata update} After the new entry is inserted into
the TLB, its \scheme{} metadata is updated to inform \scheme{} of the next 
replacement decision. First, the signature of the new entry is used to index the
prediction table and then the corresponding counter is read out
(Figure~\ref{fig:mhrpalgo}, line~\ref{op:indexsignnew}). The counter value is
thresholded and used to decide if the incoming entry should be predicted dead
or live in the future. The resulting prediction status is then used to update
the prediction bit in the \scheme{} metadata.

On a TLB hit \scheme{} updates its prediction table only if the current
access is the first hit to the TLB entry 
line~\ref{op:newset}).  These optimizations
improve both performance and energy as they reduce the frequency of
access to the \scheme{} prediction table to only 10.14\% of all TLB
accesses (Figure~\ref{fig:accessrate}). In addition, for smaller
prediction tables, they prove very effective at improving MPKI by
reducing aliasing (Figure~\ref{fig:hwrange}).
These optimizations and results are discussed in detail in 
Section~\ref{sec:results}).

A TLB hit (Figure~\ref{fig:mhrpalgo}, line~\ref{op:hit}) involves the
following operations:

\paragraph{Prediction table update}
On a hit the prediction table is accessed only on the first access or reuse
(Figure~\ref{fig:mhrpalgo}, line~\ref{op:newset}). The old signature
in the entry (Figure~\ref{fig:mhrpalgo}, line~\ref{op:indexsignhit})
is used to index the prediction table and the corresponding counter is
then decremented to assure this entry will be predicted as live under
the same conditions in the future (Figure~\ref{fig:mhrpalgo},
lines~\ref{op:updatealive} and~\ref{alg:updatetables}). Then
the old signature is replaced with the new one.

\paragraph{Prediction metadata update}
The new signature of the hitting entry is used to index the
prediction table and then the corresponding counter is read out. The counter value is
thresholded and used to decide if that entry should be predicted dead
or live in the future. The resulting prediction status is then used to update
the prediction bit in the \scheme{} metadata.
Figure~\ref{fig:mhrpalgo} summarizes the steps taken during a TLB hit. 

\subsection{Adapting Training Algorithm for TLB}
Access to a TLB reuse predictor must be fast and energy efficient, as the TLB
is on the critical path to accessing memory.  Thus, we are motivated to
minimize the number of updates made to prediction structures.  We find that
two specific events are sufficient for an accurate training update:

  \begin{compactitem}
   \item The first hit of an entry.
   \item A miss in a set with no dead entry (this leads the algorithm to choose an entry to evict based on LRU.)
   \end{compactitem}

With this technique, \scheme{} reduces the access ratio to the prediction tables by
90\% compared to SHiP and GHRP (see Figure~\ref{fig:accessrate}), which must access
tables on \textit{every} access to the TLB.

  \begin{table}[!ht]
 \small
\begin{center}

      \begin{tabular}{|l|c|} 
    
\hline {\bf Component} & {\bf Size}\\ \hline \hline
Prediction bits & $1 \textrm{ bit}\times 1024=$ 128B \\ \hline
{FirstHit bits} &{ $1 \textrm{ bit}\times 1024=$ 128B }\\ \hline

Signature bits & $16 \textrm{ bits}\times 1024=$ 2{KB} \\ \hline
Path history register&$64 \textrm{ bit}\times1=$ 8B \\ \hline
Cond. history register&$64 \textrm{ bit}\times1=$ 8B \\ \hline
Uncond. history register&$64 \textrm{ bit}\times1=$ 8B \\ \hline
Counters & 128B .. 8{KB} \\ \hline
Total & {2.775KB .. 8.265KB}\\ \hline

      \end{tabular}
\end{center}
  \caption{Storage overhead of CHiRP for a 1024 entry, 8-way L2 TLB with 4{KB} pages.}
      \label{tab:1}
    \end{table}

   \begin{table}[!ht]
   \small
\begin{center}
      \begin{tabular}{|l|c|} 
\hline {\bf Processor} & {\bf Parameter}\\ \hline \hline
L1 i-Cache & 64{KB}, 8 way, 4 cycles  \\ \hline
L1 d-Cache &  64{KB}, 8 way, 4 cycles  \\ \hline
L2 Unified Cache & 256{KB}, 16 way, 12 cycles   \\ \hline
L3 Unified Cache &  8{MB}, 16 way, 42 cycles  \\ \hline
DRAM & 240 cycles \\ \hline
Branch Predictor & Hashed perceptron, 4K \\
		& entry BTB, 20 cycle \\ 
		& miss penalty\\ \hline
L1 i-TLB 	& 64 entry, 8 way, 1 cycle \\ \hline
L1 d-TLB 	& 64 entry, 8 way, 1 cycle \\ \hline
L2 Unified TLB 	& 1024 entries, 8 way,  \\
		& 8 cycle hit latency, 20 to \\
		& 360 cycle miss penalty \\ \hline

      \end{tabular}
\end{center}
\caption{Simulation Parameters}
      \label{tab:2}
    \end{table}

\section{Methodology}\label{sec:method}

To implement and test \scheme{}, we use the simulator and traces released for
the recent Championship Value Prediction Competition (CVP1)\cite{CVP1website}.
There are hundreds of traces available (of which we use 870), coming from a 
variety of workload categories of interest to Qualcomm who provided them. In 
particular, the workloads come from the team working on their (now defunct) 
server project. The traces contain SPEC, database, crypto, scientific, web, 
``big data'' and other applications, many of which exhibit interesting address 
translation behavior.  The traces contain very detailed information such as 
instruction type, register values, effective addresses of loads and stores, and 
data values, making them suitable to drive a performance simulator.
Short traces are simulated completely, while long traces are allowed to run
for 100 million instructions. 
To measure the performance numbers we built a timing-approximate performance
model similar to previous work~\cite{Cox2017Multiple}. 

Our model simulates first-order sources of processor latency such as 
the memory hierarchy composed of L1 i-TLB and L1 d-TLB, L1 i-cache, 
L1 d-cache, L2 and L3 unified caches, DRAM, a branch prediction unit that includes 
an indirect branch predictor, a conditional branch predictor with branch target buffer, 
and an in-order pipeline model.  We use a hashed perceptron predictor as the 
branch direction predictor~\cite{HashedPerceptron}. 

We measure misses per 1000 instructions (MPKI) as well as instructions per
cycle (IPC) based on the simulated microarchitecture across a variety of page table 
walk latencies derived from previous work. A recent reverse engineering 
study on TLB~\cite{gras_translation_2018} reported a range of L2 TLB miss 
penalties for Intel microarchitecture: 230 cycles for Coffeelake, 272 cycles 
for BroadwellXeon, 212 cycles for Skylake and 18 cycles for Haswell. 
A related study on TLBs, Li {\em et al.}~\cite{Li2013PSTLB}, uses 150 cycles for L2 
TLB miss penalties. We measure speedup for a range of 20 to 340 cycles page walk 
latencies, shown in Figure~\ref{fig:speedup}.

We model static re-reference interval prediction (SRRIP), signature-based hit
prediction (SHiP), global history reuse prediction (GHRP) and control-flow
history reuse prediction (\scheme). CHiRP keeps metadata for each L2 TLB
entry. CHiRP also uses one prediction table. Each of the entries in the table
contains a two bit counter. The additional metadata for each entry consists of
1 prediction bit, 3 bits to maintain LRU positions, and 16 bits of signature.
Table-\ref{tab:1} summarizes the storage requirements for CHiRP for a
1024-entry 4KB page size L2 TLB with 8-way associativity.

We assume a 4KB page size similar to previous work~\cite{Li2013PSTLB}.  Large
pages are supported in current microarchitectures, {\em e.g.} Intel's Skylake
supports page sizes of 4KB, 2MB, 4MB, and 1GB. Large pages can reduce capacity
misses in TLBs when program behavior exhibits high locality. However, 4KB
pages are still the norm for most mobile and desktop operating systems,
providing a good balance between impact of page faults for workloads with good
locality and impact of fragmentation for workloads with poor locality. 
It would be easy to say, ``just use large pages'' but the performance of legacy
systems, mobile apps, cloud computing workloads, etc. that continue to use 4KB
pages matters to users of those systems. 

The complexity of variable-sized TLB entries (as compared to fixed-sized lines
in cache replacement) further complicates efforts to improve TLB replacement. 
Entries for different sized pages share the L2 TLB; as the L2 TLB is built 
for capacity, it is not partitioned among page sizes.

Reasoning about how to do replacement with a mix of page sizes is an
interesting problem we plan to tackle in future work; imagine, when one entry
covers 4KB and another covers 2MB, which one is is more important to keep? It is
no longer a matter of pure replacement in the sense of trying to achieve
B\'el\'ady's optimal result~\cite{belady1966study}, but now requires taking 
into account the different costs of replacing different sized 
entries\cite{albers1999page, young2002line, chrobak2012caching, berger2018practical}. 
This question is beyond the scope of this initial study. In addition, large 
pages' susceptibility to memory fragmentation requires simulating traces 
with varying levels of known fragmentation behavior, complicating an already 
complex issue.  We hope this initial work invites the community to consider 
and tackle the problem of TLB replacement further from the surface work seen 
so far.  Thus, in this initial study of predictive replacement policies for 
TLBs, we focus on the standard 4KB page size.

  \section{Results} \label{sec:results}

In this section, we describe the results of experiments simulating the \scheme{}
policy and demonstrate its superior over policies used in previous work. 
Results with a range of hardware budgets are presented. In the absence of 
public data about L2 TLB size, our calculation accounts only for tag, physical page
number, replacement metadata, protection bits, valid bit, and ASID,
estimating 118 bits for a TLB entry, giving 14.75KB for a
1024-entry TLB.
A 6\% TLB overhead places CHiRP overhead of 1KB, which still offers a 28\% MPKI improvement (see Figure {\ref{fig:hwrange}}). 

\begin{figure*}[htb!] 
  \centering
  \caption{Effect of correlating features, transforming input{, shifting PC bits and scaling}, signature formula and 
  prediction table update policies on reducing misses in the L2 TLB. 
  The $x$-axis is the reduction rate of average MPKI over 870 traces over a baseline LRU. Previous predictive replacement policies need specific optimizations to work for L2 TLB.  Results show the advantage of \scheme{}.}
  \includegraphics[width=\textwidth]{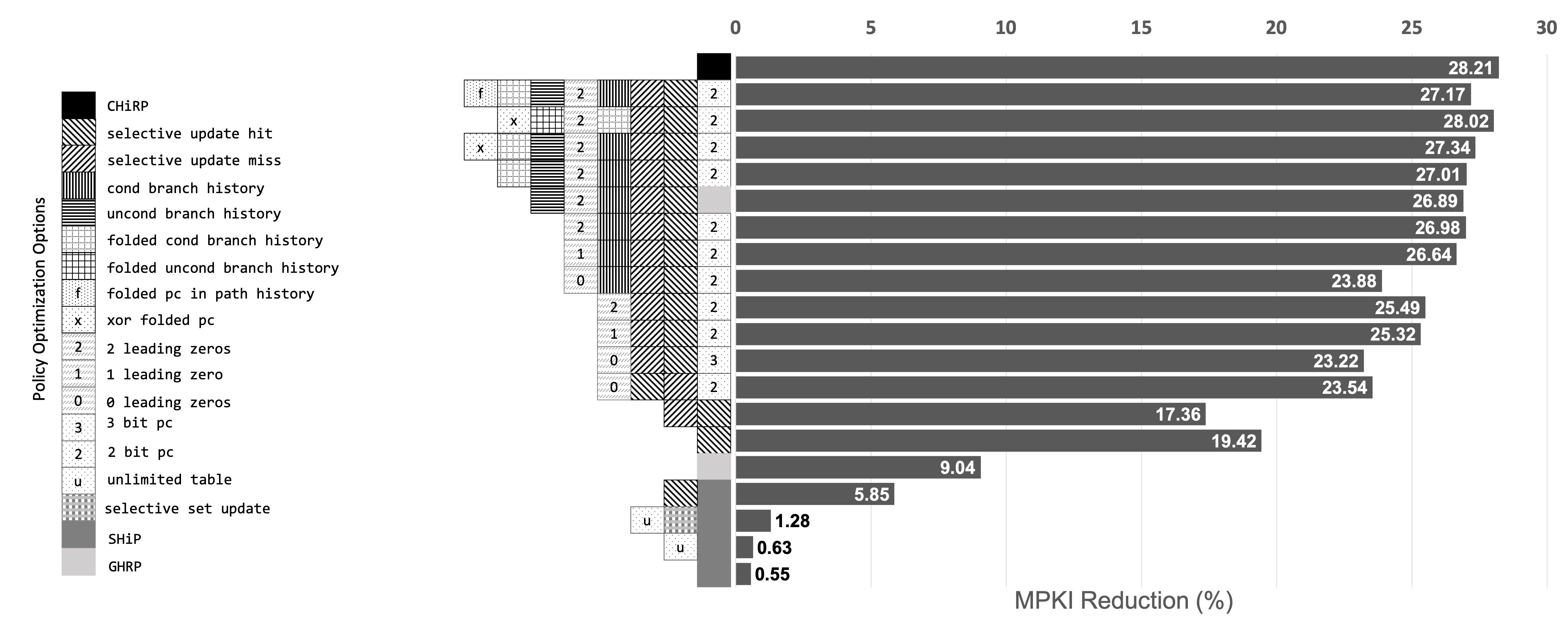}
  \label{fig:motive}
\end{figure*}
\subsection{MPKI Results}
  
\begin{figure*}[htbp!] 
\centering
\includegraphics[width=\textwidth]
{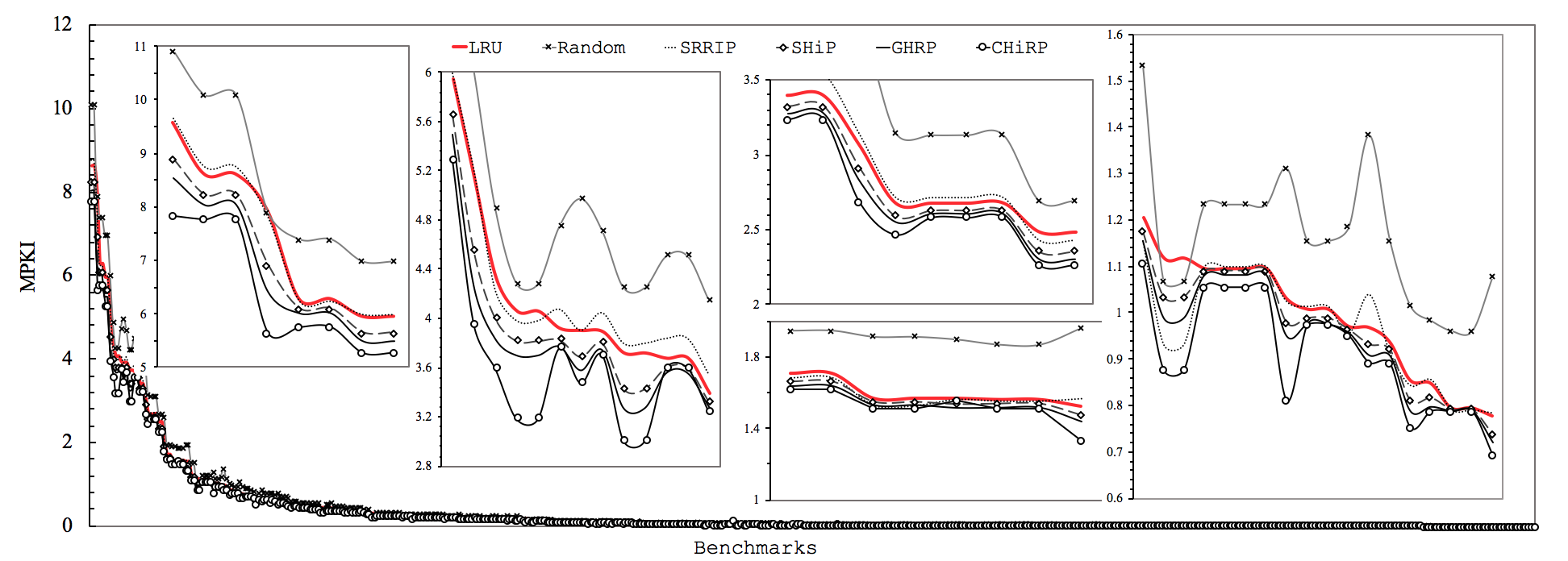}
\caption{MPKI comparison of various policies. The horizontal axis shows the benchmarks 
    in the order of sorted MPKI for LRU. Multiple zoomed-in areas of the graph are 
    shown in insets.}
\label{fig:scurvetlb}
\end{figure*}

Figure~\ref{fig:scurvetlb} shows an S-curve of MPKIs for 870 benchmarks.
The $x$-axis shows the benchmarks in order of sorted MPKI for LRU and is
compared with other policies. Insets highlight key areas of the graph.

LRU and Random yield an average 1.51 and 1.47 MPKI, respectively. 
SRRIP, which uses a simple static prediction on each TLB
entry placement and has a lower cost than LRU, yields 1.35 MPKI, a 10.36\%
improvement over LRU. SHiP, a PC-based reuse predictor, gives an
average 1.50 MPKI, performing almost the same as LRU with 
0.88\% improvement. GHRP, which uses a more
detailed prediction signature, yields an average 1.37 MPKI, or a 9.03\%
reduction in misses over LRU. \scheme{}, with a signature specially designed for
TLB replacement, gives an average MPKI of 1.08\% an improvement of 28.21\% over
LRU.  Nearly all of the tested benchmarks exhibit considerable MPKI reduction
under \scheme{}, achieving an improvement of 58.93\% in some cases.

These results demonstrate that the case for LRU as a TLB replacement policy is weak, 
as even Random replacement slightly outperforms it. SRRIP, which is a simpler and
low-overhead policy, could more conveniently be deployed in current
processors, yielding better performance. \scheme{} is somewhat more complex but
yields the best improvement, more than double the improvement provided by SRRIP.

\subsection{Accesses to Prediction Table} Predictive replacement policies
often access tables of counters to make a prediction. Previous work on cache
and BTB replacement policies read out from the tables on every access to the
cache/BTB to make a prediction for the next access. The tables are also
modified frequently as counters are updated.

\scheme{} only accesses the prediction table on a TLB miss or on a hit
to a TLB set different than the one accessed last, following our
selective hit update policy. It follows from this that consecutive
hits to the same set do not result in writing or reading from the
prediction table but only updates to the signature in the TLB
entry. Updating the signature bits in the TLB entry has the same
overhead as an LRU stack update.  Thus, the energy and timing
properties of \scheme{} are considerably more favorable to
implementation than techniques based on previous predictive
replacement policies.

Figure~\ref{fig:accessrate} shows a density plot of the rate of the number of
accesses to prediction table over accesses to the TLB for SHiP, GHRP and \scheme{}. 
The plot shows the distribution of access rates over all the benchmarks. For SHiP
and GHRP, the access rate has a very high variance, reaching over 100\% in
many cases. The rate can exceed 100\% because, for every TLB access, there
could be two accesses to the prediction table: one to read out the prediction,
and another to update the table for training.  For \scheme{} the access rate is
quite low, and has low variance, making for a far more practical policy for
implementation.  On average \scheme{} accesses the prediction table for 10.14\% of
the accesses to L2 TLB.

\subsection{Speedup}

\begin{figure}[ht!] 
  \centering
  \includegraphics[width=0.44\textwidth]{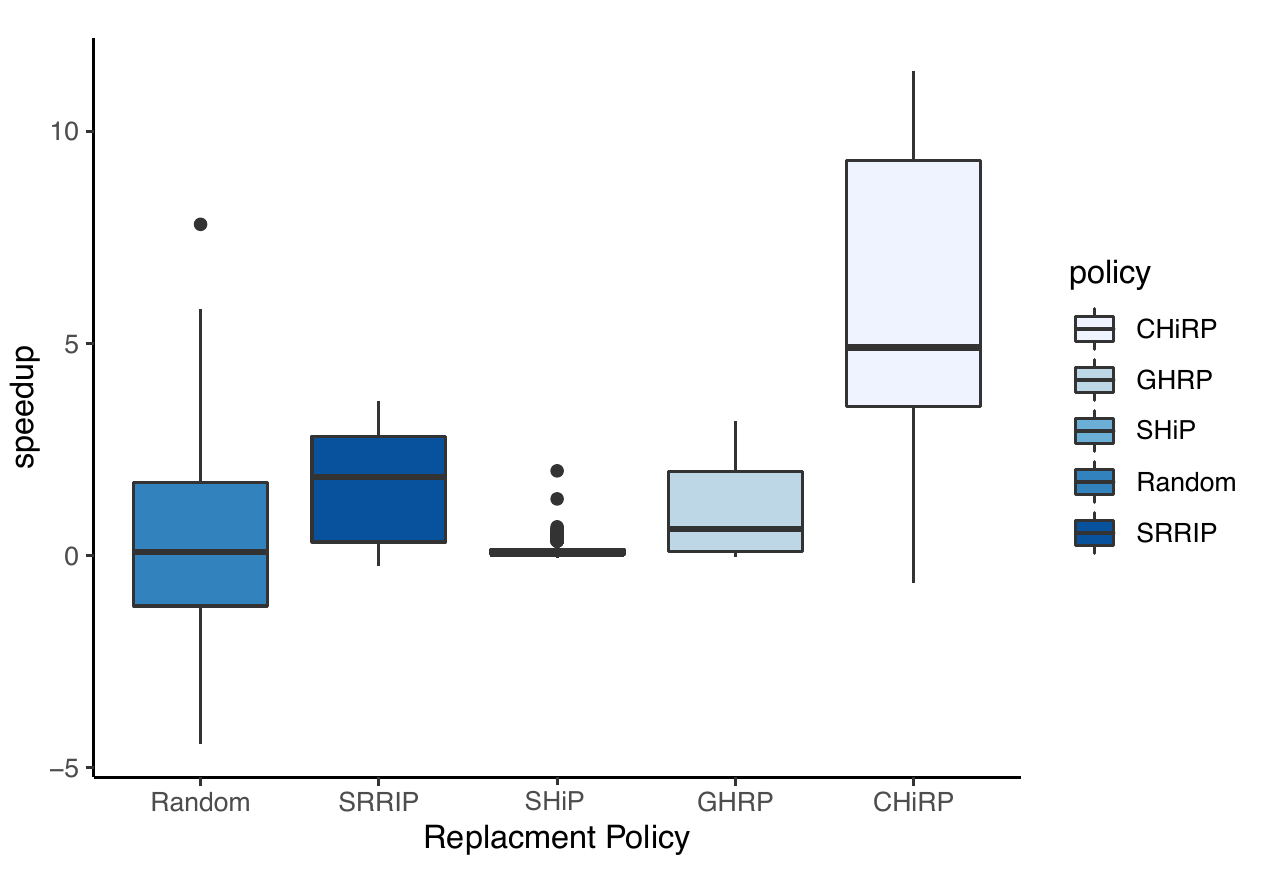}
  \caption{Speedup for 870 traces.}
  \label{fig:speedup2}
\end{figure}

Figure~\ref{fig:speedup2} shows speedup for various policies with TLB miss penalty of 
150 cycles. Page walk latency depends on several microarchitectural and software parameters.
Thus, we explore a range of L2 TLB miss penalties to provide an estimate of
performance under different assumptions. With TLB miss penalty of 150 cycles,
\scheme{} improves performance by 4.80\% compared to 0.42\% for Random, 1.65\% for 
SRRIP, 0.13\% for SHiP, and 0.94\% for GHRP. At higher latencies, the advantage of 
predictive policies grows. With a penalty of 320 cycles, representing more memory 
intensive behaviors, \scheme{} provides a speedup of over 10\%. Other predictive 
replacement policies do not provide significant speedup. Clearly, \scheme{} 
provides significant improvement to performance over all other replacement policies.
  
\subsection{TLB Efficiency}

Cache efficiency is the average amount of time in which a block was live in
the cache.  We calculate cache efficiency~\cite{bur95} for TLB entries instead
of cache blocks.  Figure~\ref{fig:heatmap} depicts cache efficiency for the L2 TLB 
for 870 benchmarks. Each row is the cache efficiency of one benchmark for various
policies scaled by LRU. Benchmarks are sorted from low to high cache
efficiency from down to up respectively. Figure~\ref{fig:heatmap} shows Random helps
improve the cache efficiency of some benchmarks (most which had high efficiency
already) but \scheme{} removes most of the dead entries in TLB for all 870
benchmarks. \scheme{} improves average cache efficiency over 870 traces by 8.07\%
compared to LRU. This number is 2.92\% for GHRP, 1.85\% for SHiP, 2.84\% for SRRIP, 
and 3.10\% for Random. {Thus, the MPKI improvement 
in} {\scheme{}} {comes from reducing dead entries and increasing live entries in the TLB.}

\subsection{Complexity and Efficiency}

Note that \scheme{} is more complex than simple replacement policies
such as LRU and RRIP. However, it is far less complex than, for example,
branch prediction techniques such as TAGE~\cite{TAGEHPCA} and
perceptron~\cite{PerceptronPredictor} that have been implemented in recent
processors. These predictors require far more logic, dynamic energy, and state
than \scheme{} and have tighter timing constraints. Thus, we believe the complexity
of \scheme{} is very manageable given its benefits to front-end performance.

Consistent with branch predictor implementation,
\scheme{} only updates the tables of counters at commit with right-path branches to 
prevent pollution of the tables. For misprediction recovery, \scheme{} maintains two path 
histories: the speculative history updated using the outcome of the branch predictor, 
and a non-speculative history updated when a branch commits. 

The energy overhead of predictive policies results from accesses to 
the prediction table {and updating respective metadata in the TLB entry}.
Because \scheme{} reduces the number of accesses to the 
prediction table by 90\% compared to previous predictive policies ({\em e.g.} SHiP and 
GHRP), energy consumption {related to accessing the tables} should be 
less of a concern (Figure~\ref{fig:scurvetlb}).

{\scheme{}} {requires more accesses to the prediction metadata 
in the TLB compared to SHiP. While SHiP only updates metadata at insertion 
time, CHiRP updates the signature in the TLB during both hit and insertion.}{
\scheme{}}{updates the metadata in the same manner as updating LRU bits. While 
SHiP cannot perform better than LRU for the L2 TLB, updating the metadata on every 
access is the cost for an accurate predictor.}

\begin{figure}[htb!] 
\centering
\includegraphics[width=2.5in,height=2.5in]{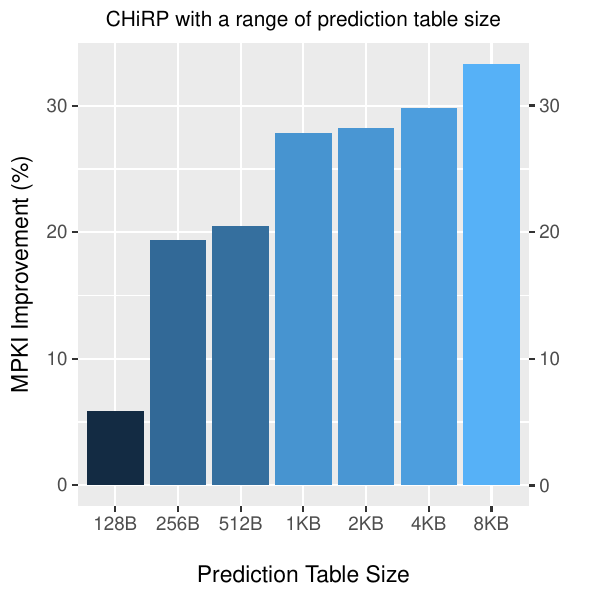}
  \caption{MPKI improvement over LRU for \scheme{} with a range of prediction 
    table sizes.}
  \label{fig:hwrange}
\end{figure}

\subsection{Impact of Predictor Size}

Figure~\ref{fig:hwrange} shows MPKI improvement over LRU for \scheme{}
with a range of prediction table sizes. The size of the prediction
table has an impact on area, energy, and timing, so we would like to
choose a size that yields a good improvement while maintaining a
reasonable cost. At a very small hardware budget of 128B, we note that
even though we may experience higher conflicts rate in the prediction
table, \scheme{} still yields up to 7\% MPKI improvement over LRU. As
we double the prediction table size to 256B and 512B, the MPKI
improvement increases to up to 20\% and 22\%, respectively. What this
shows is that even with a small hardware budget size of 256B,
\scheme{} doubles the MPKI improvement realized by an 8K GHRP (9\%). 
Table sizes of 1K and 2K yield similar improvement: about 28\% MPKI 
reduction; our main results are presented with a 1K budget. Gains realized 
by larger table sizes are higher, but come with larger area overhead.

\subsection{Performance Gain}

This first study focuses on single core. IPC gains in each generation are usually 
within 10-15\%, of which 20-40\% might come from the front-end (2-4\% overall), 
which is considered aggressive. A 4.8\% improvement over LRU is a significant 
milestone in this case. Figure 11 shows the speedup for \scheme{} is statistically 
significant over 870 workloads assuming a TLB miss penalty of 150 cycles. 
 
\subsection{Area Overhead vs. Performance}

\scheme{} reduces hardware overhead by two-thirds compared to GHRP
because the signature formula enables \scheme{} to use one prediction
table rather than the three needed by GHRP. The predictor cost was
evaluated for a range of extra overhead. Figure~\ref{fig:hwrange}
shows even a small 256B predictor leads to a 20\% MPKI reduction.  As
a matter of comparison, a recent study from
Intel~\cite{branchprediction2019MICRO} demonstrates a branch
prediction technique that costs 64KB hardware overhead improves IPC by 2.7\%. \scheme{} 
for a TLB with 1KB overhead and 4.80\% speedup is 13$\times$ more efficient in terms 
of speedup-per-KB overhead. This is due to the high TLB page walk latency compared to 
other miss penalties in the pipeline.

\begin{figure}[ht!] 
\centering
\includegraphics[width=2.8in,height=2.8in]{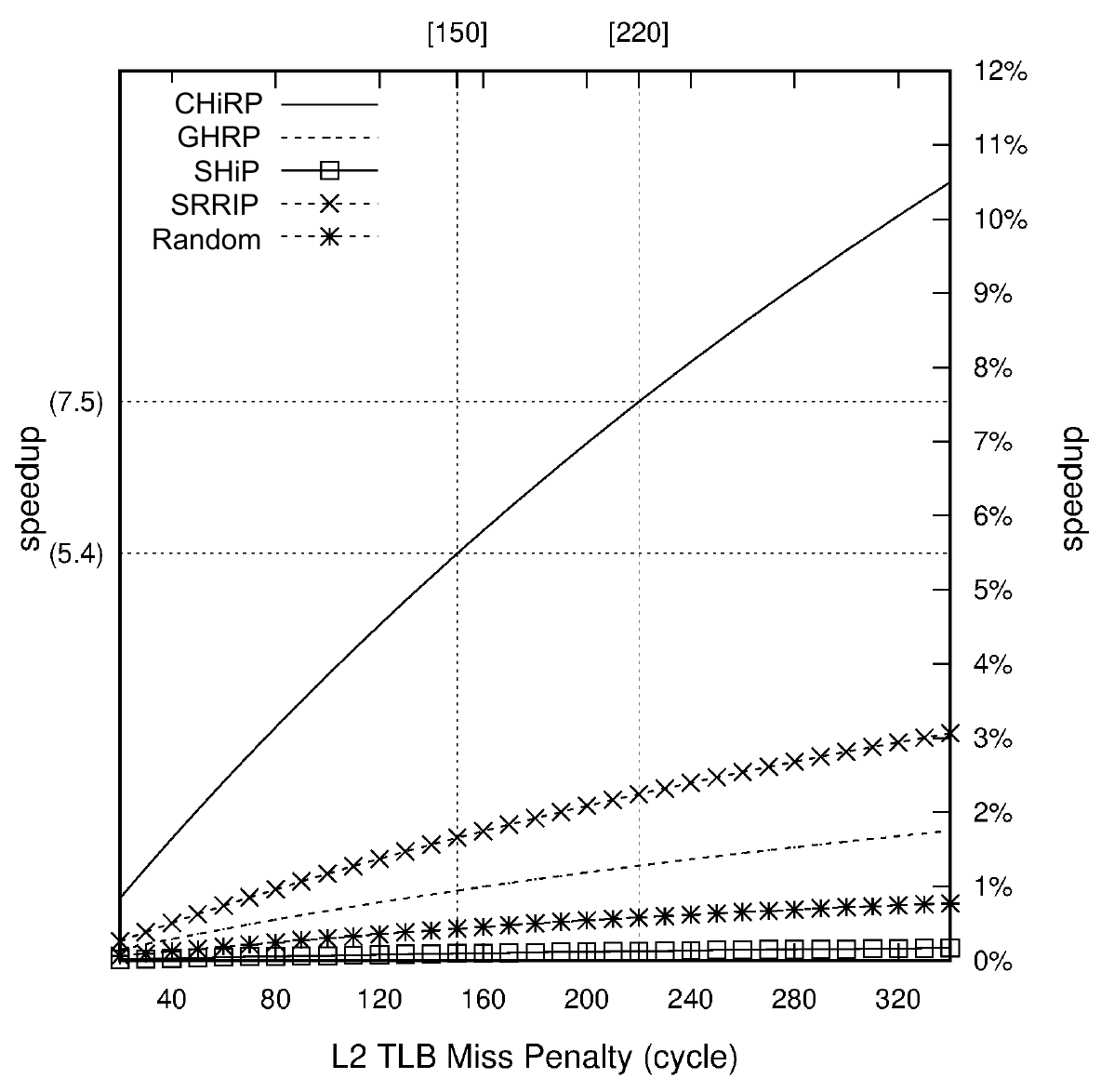}
 \caption{Average Speedup for a range of L2 TLB miss penalties over 870 traces.}
 \label{fig:speedup}
\end{figure}

\begin{figure}[ht!] 
\centering
\includegraphics[width=0.44\textwidth]{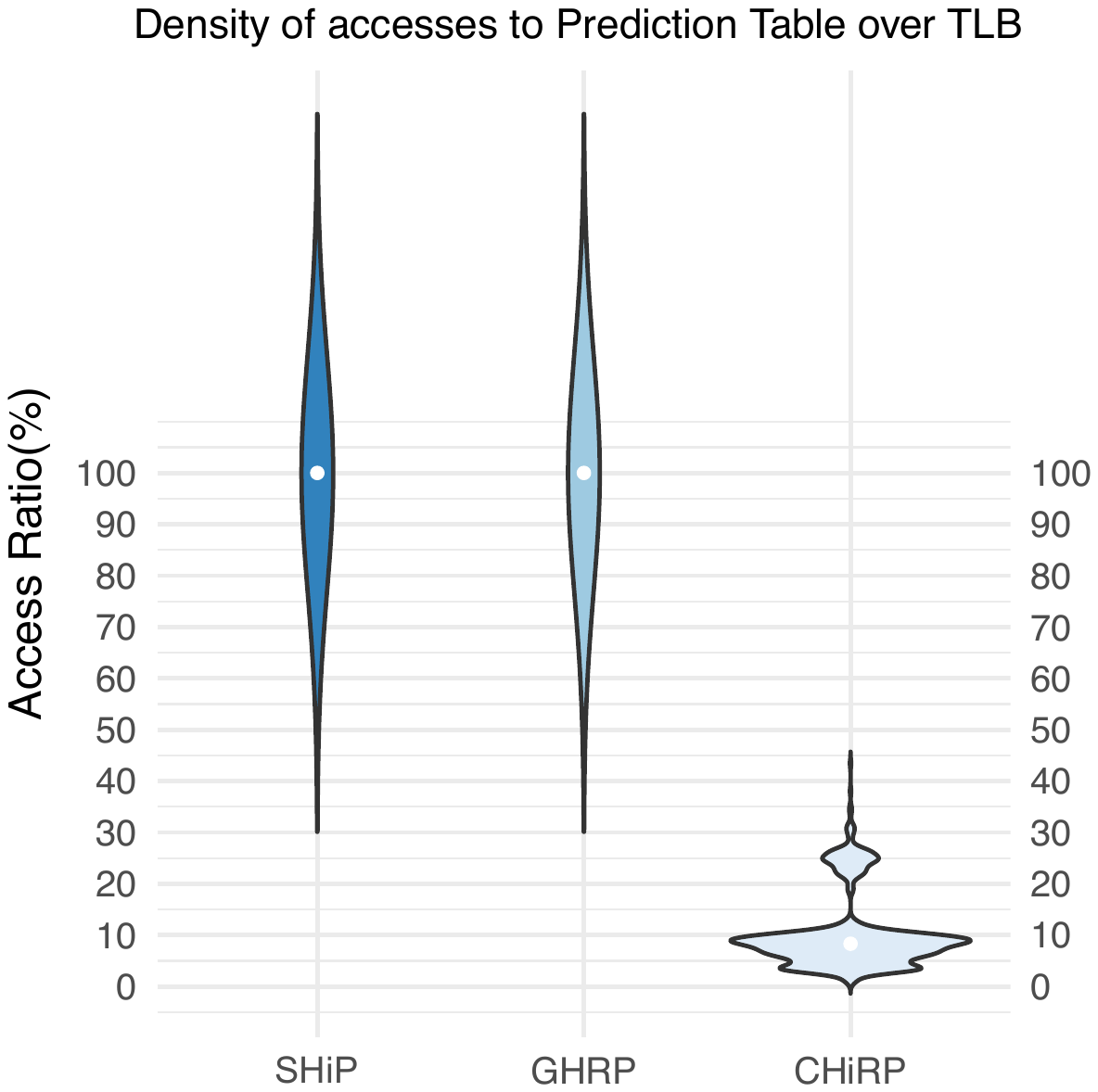}
  \caption{Density plot for rate of accesses to prediction table over 
    accesses to TLB for SHiP, GHRP and \scheme{} with mean values.}
\label{fig:accessrate}
\end{figure}

  \section{Related Work}

\subsection{Reuse Prediction}
This paper proposes reuse prediction for TLB replacement.
However, reuse prediction has a long history in the literature.  Caches often 
retain {\em dead blocks}, {\em i.e.} blocks in the cache that will not be used 
again until they are evicted~\cite{Lai2001}. Dead blocks waste space and energy in 
the cache. Lai {\em et al.} initially proposed dead block prediction~\cite{Lai2001} 
to prefetch data into predicted dead blocks. Kharbutli {\em et al.} propose a 
counter-based dead block prediction approach~\cite{kha08} for replacement and 
bypass. Liu {\em et al.}~\cite{liu08} propose a predictor leveraging the burst-like 
nature of accesses to the L1 cache. Teran {\em et al.} propose using perceptron 
learning for reuse prediction~\cite{PerceptronReuse,MultiperspectiveReuse}.

Dead block prediction has been evaluated in the context of making replacement decisions 
in the L1 data cache~\cite{Lai2001,liu08,SamiraMICRO,SHiP,PerceptronReuse,
MultiperspectiveReuse}, last-level cache~\cite{SamiraMICRO,liu08,VVCCMP},
prefetching~\cite{Lai2001,hu02}, bypassing~\cite{chi1989improving,wu2002compiler,
johnson1999run,tam1999active,johnson1997run,rivers1998utilizing,rivers1996reducing,
milutinovic1996split,gonzalez2014data,tyson1995modified,li2002leakage,dybdahl2006enhancing,
jalminger2003novel}, power reduction~\cite{abella2005iatac,kax01}, and cache 
coherence protocol optimization~\cite{lai00,lebeck1995dynamic,som04}. However, 
no replacement policy has been proposed for the TLB based on dead block prediction.

\subsection{Translation Lookaside Buffers}

Previous work shows that superpages, {\em i.e.} any page size larger than the default, 
can increase TLB reach and reduce TLB misses. Large pages are especially beneficial 
for applications with a hard upper bound of memory usage in terms of maximal heap 
size~\cite{Talluri1992Tradeoffs, Talluri1994Surpassing, Ganapathy1998General,
Swanson1998Increasing, Fang2001Reevaluating, Peng2006JavaApps, Kandiraju2002signmetric}.
Still, Peng {\em et al.}~\cite{Peng2006JavaApps} show that while superpages can remove nearly
all the TLB miss overhead of some benchmarks, an increased page size of 1MB cannot cover the
working set of some benchmarks due to unpredictable memory access patterns. If memory access 
patterns are predictable, TLB misses can be reduced through prefetching and
speculation~\cite{Saulsbury2000Recency, Kandiraju2002Going, Bhattacharjee2009Characterizing,
Bhattacharjee2017Triggered, Barr2011SpecTLB}.

Conversely, using superpages may unnecessarily increase the memory footprint of an
application, resulting in elevated, but useless, paging traffic and memory allocation.
Additionally, handling multiple page sizes increases complexity in the operating
system~\cite{Talluri1992Tradeoffs, Talluri1994Surpassing, Romer1995Reducing, 
Ganapathy1998General, Navarro2002Practical, Arcangeli2011THP}. Algorithms to evaluate the 
need for larger pages based on applications' behavior are essential for choosing the
appropriate page size. Techniques for mapping multiple smaller pages into a single 
superpage TLB entry~\cite{Talluri1994Surpassing, Pham2012CoLT, Pham2014Increasing,
Park2017Hybrid, Cox2017Multiple} reduce splintering and make superpage usage more 
efficient, but require deep OS-hardware co-design.

With the prevalence of chip multiprocessors (CMPs) and parallel workloads, recent TLB 
work has focused on distributed TLBs in architectures. Cooperative
TLB~\cite{Bhattacharjee2010Cooperative,Srikanthaiah2010Synergistic} and shared last-level
TLB~\cite{Bhattacharjee2009Characterizing, Bhattacharjee2011Shared,Bharadwaj2018Scalable} schemes have been proposed.

  \section{Conclusions and Future Work} \label{sec:future}

This paper extensively investigated the replacement policy of TLBs, which has
been rarely studied in previous work.
 In the past, the only way to provide a predictive policy for larger cache structures was to use a sampling method. We show that sampling does not work in the L2 TLB.
 The idea of sampling is to generalize learning over sets; we used the granularity of L2 TLB entries to generalize learning instead of sampling. Prior work does not recognize the effect of the granularity of a structure on sampling and dead block prediction. The signatures of previous policies do not detect dead blocks in the L2 TLB. Because they do not follow control flow, it was impossible for them to learn the reuse patterns in the L2 TLB properly. They end up averaging over traces, whereas we present a specific signature that tracks the trace of dead blocks in a large granularity environment while minimizing the prediction counters’ fluctuations. That allows CHiRP to use a small table with fast convergence, providing a predictive replacement policy that fits into constraints of the L2 TLB for the first time. In future work we plan to extend \scheme{} 
to TLBs with mixed page sizes.

  \section{Acknowledgement}
We thank Jeffrey N. Collins and Andrew Worthen for their help and comments during the drafting of
this paper. This research was supported by NSF grants CCF-1912617, CNS-1938064, and CCF-1332598 as well as generous gifts from Intel Labs.

\end{spacing}

%

%


%

\end{document}